\documentclass{aa}
\usepackage{graphicx}
\usepackage{txfonts}
\begin{document}
   \title{Full computation of massive AGB evolution. I. 
          The large impact of convection on nucleosynthesis}

   \author{P. Ventura
          \and
          F. D'Antona
          }

   \offprints{P. Ventura}

   \institute{Osservatorio Astronomico di Roma
              Via Frascati 33 00040 MontePorzio Catone - Italy\\
              \email{ventura@mporzio.astro.it, dantona@mporzio.astro.it}
             }

   \date{Received ... ; accepted ...}

   \abstract{
   It is well appreciated that the description of overadiabatic convection
   affects the structure of the envelopes of luminous asymptotic giant
   branch (AGB) stars in the
   phase of ``hot bottom burning '' (HBB). We stress that this important
   uncertainty in the modeling plays a role which is much more dramatic
   than the role which can be ascribed, e.g., to the uncertainty in the
   nuclear cross-sections. Due to the role tentatively attributed today to 
   the HBB nucleosynthesis as the site of self-enrichment of Globular
   Clusters stars, it is necessary to explore the difference in 
   nucleosynthesis obtained by different prescriptions for convection.
   We present results of detailed evolutionary calculations 
   of the evolution of stars of intermediate mass during the
   AGB phase for the metallicity typical of the
   Globular Clusters that show the largest spread in CNO abundances
   ($Z\sim 10^{-3}$). We follow carefully the nucleosynthesis at 
   the base of the external convective region,
   showing that very different results can be obtained according to
   the presciption adopted to find out the temperature gradient within
   the instability regions. We discuss the uncertainties in 
   the yields of the various chemical species and the role which these
   sources can play as polluters of the interstellar medium.

   }

   \maketitle
%

\section{Introduction}

The intermediate mass stars (i.e. stars with initial masses 
$1 \leq M/M_{\odot} \leq 7$, hereinafter IMS) experience 
a phase of thermal pulses (TPs) shortly after the exhaustion 
of central helium (Schwarzschild \& Harm 1965, 1967;
Iben 1975, 1976). 
A CNO-burning shell supplies the global nuclear energy 
for $\sim 90\%$ of the time; 
periodically, a thermal instability (associated with the 
narrow dimensions of the helium-rich layer above the CO 
degenerate core and with the steep dependence of the 
cross-section of $3\alpha$ reactions on the temperature)
triggers a runaway-like ignition of helium burning, which 
temporarily extinguishes the H-burning shell. As the mass
of the CO core increases, the CNO-burning shell becomes hotter
and hotter, so that the luminosity of the star increases: on
the HR diagram the evolutionary tracks climb the so called
"Asymptotic Giant Branch" (hence the term "AGB evolution",
commonly adopted). During their evolution these structures
suffer strong mass losses, which eventually stop their
AGB evolution, leaving just a central remnant which later
evolves into a CO white dwarf.

In the last decades the interest in the detailed 
evolution of these stars
has grown, also because they have been suggested as 
pollutors of the medium during the early phases
of evolution of globular clusters 
(GCs), and responsible for the chemical anomalies 
(in terms, e.g.,  of oxygen-sodium and 
magnesium-aluminium anticorrelations) which are 
observed in giants and turn-off stars of GCs.
At least in the latter because of the low temperatures 
involved the chemical anomalies cannot be due
to ``in-situ'' processes. The idea 
behind this hypothesis (to which we usually
refer as ``self-enrichment'') is that
the base of the external envelope of the 
most massive IMS 
during the AGB evolution may become so hot 
($T_{\rm bce} > 30\cdot 10^6$K) that it triggers strong 
nucleosynthesis; the consequent changes in the originary 
chemical abundances might easily reach the surface of the 
star because of the rapidity of the convective motions, so that
the ejecta of these stars are contaminated by such
an advanced nucleosynthesis.  The velocity of the ejected
material should be sufficiently low to survive ejection
from the cluster (D'Antona et al. 1983; Ventura et al.
2001, 2002).

The self-pollution scenario is not without its shortcomings,
as pointed out by Denissenkov \& Herwig (2003): the 
temperatures needed to deplete oxygen would also
lead to sodium destruction, and to magnesium isotopic
ratios $^{25}\rm{Mg}/^{24}\rm{Mg}$ and 
$^{26}\rm{Mg}/^{24}\rm{Mg}$ much larger than observed.

Unfortunately, the theoretical models for the 
AGB evolution are characterized by several uncertainties 
associated with the input physics which is commonly 
adopted in the stellar evolutionary codes: 
in the last decade the attention of modelists has 
been mainly focused on: a) the amount of extra-mixing 
from the base of the external region (Herwig 2000, 2004), 
which favours a 
larger efficiency of the so called third dredge-up 
phenomenon (i.e. the external convective region which
reaches stellar layers previously touched by efficient
$3\alpha$ burning after a thermal pulse); b) the rate
of mass loss suffered by these stars; c) the nuclear
network adopted, particularly concerning the
nucleosynthesis of Neon, Sodium and Magnesium isotopes. 

The present work does not discuss
the plausibility of the self-enrichment scenario, but
is aimed at pointing out quantitatively that,
before going into the details of the chemical composition 
of the ejecta of AGBs, it is mandatory to understand
to which extent the results obtained depend on the
assumptions made in calculating the models.

We focus our attention on one of the most
relevant uncertainties connected with stellar evolution,
i.e. the treatment of convection, whose influence upon 
AGB evolution in the context of HBB is well documented
in the literature, starting with the classic Renzini \& Voli
(1981) paper, and with the discussion by Sackmann \& 
Boothroyd (1991) and Bl\"ocker \& Sch\"onberner (1991). 
Finally, D'Antona \& Mazzitelli 
(1996) discussed that the ``Full Spectrum of Turbulence''
(FST, Canuto \& Mazzitelli 1991) model for convection leads 
to a much more efficient HBB than the Mixing Length Theory
(MLT). Neverthless, in these last years, no mention 
has been made in many recent detailed computations 
of the description adopted for convection, and, 
more importantly, of the resulting uncertainty 
in the nucleosynthesis.
 
We will compare the results obtained 
with the two local models currently available, i.e.
MLT and the FST convective model; within the MLT
framework, we discuss the role played by the free 
parameter $\alpha$. 

We show that the main physical properties of the models
depend strongly on the efficiency of convection, and 
discuss the implications for the self-enrichment
scenario. 

\section{The evolutionary code}
The stellar evolution models discussed 
in this paper were calculated by 
the code ATON, a full description of which can be found in 
Ventura et al. (1998) (ATON2.0 version). 
In this paper the interested reader may find 
a detailed description of the numerical 
structure of the code, and of the macro- 
and micro-physics used to simulate the stellar 
evolutions. 

The code has now been updated concerning the nuclear 
network, which has been widened in order to include more 
chemical elements and nuclear reactions, according to 
the detailed description given below. The current version is 
therefore ATON2.1.

Here we briefly recall the most important input 
data adopted.

\subsection{The convective model}
The code allows us to calculate the temperature gradient
within instability regions either by adopting the
traditional MLT (Vitense 1953; B\"ohm-Vitense 1958), 
or the FST model (Canuto \& Mazzitelli 1991; 
Canuto et al. 1996). The interested reader may 
find a detailed 
description of the physical differences between 
the two models in Canuto \& Mazzitelli (1991).

Briefly, we recall that within the MLT scheme both the
dimensions of the convective eddies and the mixing length
are assumed to be directly proportional to the local
value of $H_p$, the pressure scale height
($l=\alpha H_p$); the free parameter $\alpha$
is calculated in order to reproduce the evolution of the 
Sun. The most recent estimates give $\alpha=1.7$. In the
FST model the mean dimension of the convective eddies 
is found by integrating over the whole spectrum of the
dimensions, and the mixing length is simply assumed
to be $l=z$, the distance from the closest convective
boundary.

The differences between the results provided by the two
models can be summarized as follows: 
\begin{itemize}

\item{
within high-efficiency convective regions (i.e. 
stellar regions where most of the energy is carried by 
convection) the level of overadiabaticity 
$\nabla-\nabla_{\rm ad}$, 
(where $\nabla={dlogT\over dlogP}$) 
required by the FST models is lower}

\item{within low-efficiency instability regions (e.g.
low density convective zones) the FST model exhibits
a superadiabaticity peak which is not found within
the MLT framework}

\end{itemize}

\subsection{Nuclear network}
With respect to Ventura et al. (1998) the nuclear network has been 
widened and now includes 30 chemicals:$rm{H}$,$\rm{D}$,
$^3\rm{He}$,$^4\rm{He}$,$^7\rm{Be}$,
$^7\rm{Li}$,$^{12}\rm{C}$,$^{13}\rm{C}$,$^{14}\rm{N}$,
$^{15}\rm{N}$,$^{16}\rm{O}$,$^{17}\rm{O}$,
$^{18}\rm{O}$,$^{18}\rm{F}$,$^{19}\rm{F}$,$^{20}\rm{Ne}$,
$^{21}\rm{Ne}$,$^{22}\rm{Ne}$,$^{22}\rm{Na}$,$^{23}\rm{Na}$,
$^{24}\rm{Mg}$,$^{25}\rm{Mg}$,$^{26}\rm{Mg}$,
$^{26}\rm{Al}$,$^{27}\rm{Al}$,$^{28}\rm{Si}$,$^{29}\rm{Si}$,
$^{30}\rm{Si}$,$^{31}\rm{P}$,$\rm{n}$. 
The nuclear reactions considered are 64. They include 
all the main reactions of the p-p, CNO, Ne-Na and Mg-Al chains, 
and the $\alpha$ captures of all nuclei up to $^{26}Mg$.
The relevant cross-sections can be taken 
either from Caughlan \& Fowler (1988) or from Angulo 
et al (1999).

\subsection{Convective mixing and nuclear burning}
Since we expect a non-negligible fraction of the global
nuclear release to be generated within the convectively
unstable external regions, we decided to adopt for all the
evolution models presented here a diffusive algorithm
to deal with nuclear burning within convective regions,
in which nuclear burning and mixing of chemicals are
coupled self-consistently.
We therefore solve for each element the diffusion equation 
(Cloutman \& Eoll 1976):

\begin{equation}
$$
  \left( {dX_i\over dt} \right)=\left( {\partial X_i\over \partial t}\right)_{\rm nucl}+
  {\partial \over \partial m_r}\left[ (4\pi r^2 \rho)^2 D {\partial X_i
  \over \partial m_r}\right]  \label{diffeq}
$$
\end{equation}

\noindent
stating mass conservation of element $i$. The diffusion coefficient $D$ is
taken as

\begin{equation}
$$
D={1\over 3}ul
$$
\end{equation}
where $u$ is the convective velocity and $l$ is the convective scale 
length.

Within this diffusive framework it is necessary to specify the way 
with which convective velocities decay outside the convective 
boundaries (Deng at al. 1996a,b; Herwig et al. 1997; 
Ventura et al. 1998). 
In agreement with Xiong (1985) and Grossman (1996)  and supported 
by the numerical simulations by Freytag et al. (1996), 
we assume that convective velocities decay exponentially 
outside the formal convective boundary as:

\begin{equation}
$$
u=u_b exp \pm \left( {1\over \zeta f_{\rm thick}}ln{P\over P_b}\right) 
$$
\end{equation}

\noindent
where $u_b$ and $P_b$ are, respectively, turbulent velocity and 
pressure at the convective boundary, P is the local pressure, $\zeta$ 
a free parameter connected with the e-folding distance of the decay, 
and $f_{\rm thick}$ is the thickness of the convective regions in 
fractions of $H_p$.

\section{Early evolutionary phases}
We computed three evolutions starting with 
initial masses $M=5M_{\odot}$.
With the exception of the convective model, all the physical and
chemical input data adopted are the same, i.e.:
\begin{itemize}
\item{
Since our purpose is to test how the uncertainties connected 
with convection may reflect on the plausibility of the  
``self-enrichment'' scenario, we adopt a chemistry which is 
typical of those GCs which show most of self-enrichment like M13,
NGC6752 (Gratton et al. 2001, Sneden et al. 2004), 
i.e. $Z=0.001$ and $Y=0.24$. For all chemicals
other than helium we adopted solar-scaled abundances.
}

\item{We assumed extra mixing from the external border of
any convective region as given by a parameter $\zeta=0.02$
for the exponential decay of velocities starting from the formal 
convective border; this is in agreement with the 
calibration given in Ventura et al. (1998). 
Since our purpose is to discuss the
uncertainties connected with the treatment of convection,
no inwards overshooting from the base of the 
external convective region was adopted.
}

\item{For mass loss, Bl\"ocker's (1995) prescription was used 
in all cases: the free parameter $\eta_R$ entering  
Bl\"ocker's formula was set to $\eta_R=0.02$, in agreement with
the calibration for mass loss during the AGB evolution
given in Ventura et al. (2000).
} 

\item{The Angulo et al. (1999) cross-sections were used.}

\item{
Two models were calculated with an MLT treatment of convection
with parameters $\alpha=1.7$ (MLT17 model) and $\alpha=2.1$ 
(MLT21 model). We also calculated an evolution model 
with the FST convective prescription (FST model).
}

\end{itemize}

\begin{figure}
\caption{The tracks in the HR diagram of three models
 with initial mass $5M_{\odot}$ calculated with three
 different tretaments of convection. Full: FST models;
 dotted: MLT model with $\alpha=1.7$; dashed: MLT 
 model with $\alpha=2.1$}
         \label{HR}%
\end{figure}

Fig.~\ref{HR} shows the tracks in the HR diagram corresponding to 
the three evolution models discussed above. 
We note that the tracks are
very similar during the two major phases of nuclear burning within
the central regions, because convection within burning cores is
so efficient that the gradient is practically adiabatic, 
independently
of the convective model. The only differences can be noticed
in the colors of the red giant branch (RGB) and the early AGB
evolution. These are the regions of the HR diagram where the
stars develop a very extended convective envelope, 
in which a great part of the energy is carried by 
radiation because of the low convective
efficiency: in these cases the temperature gradient 
from the CNO-burning shell to the surface 
(hence the effective temperature of the star) depends 
sensibly on the convective model adopted. 
A full discussion of the dependency on the 
convective model of the effective temperature of IMS 
during various evolutionary phases can
be found in Ventura \& Castellani (2004).

What is most interesting here is that the main physical 
properties and the duration of the various evolutionary 
phases are approximately the same for the three models, 
because in any case the nuclear sources
are either well within regions of radiative stability, 
or inside central cores where the convective efficiency 
is extremely high, making the gradient almost adiabatic.

The phase of H-burning lasts 90Myr; during it the star develops a
central convective core of $\sim 1.4M_{\odot}$, which progressively
shrinks in mass until H-burning is extinguished in the central 
regions.   
Soon after hydrogen exhaustion the star expands, and surface
convection reaches inner layers which were previously at least
partly touched by nuclear burning (first dredge-up). The 
lowest point (in mass) reached by the base of the
external envelope is for all the three models 
$\sim 2.1M_{\odot}$ away from the centre; the whole duration 
of this phase is $\sim 70,000$ yr. 

Shortly afterwards the central core becomes hot enough
to ignite $3\alpha$ reactions. During helium burning the
star develops a central convective core of 
$\sim 0.5M_{\odot}$; the total duration of this phase
for the three models discussed is $\sim 12$Myr. 

After the central helium exhaustion, $3\alpha$ reactions
operate in an intermediate layer above the core of carbon
and oxygen. All the stellar layers above this He-burning
shell expand, and the CNO-burning shell is temporarily
extinguished. Convection penetrates inwards, reaching a
layer which is $0.9M_{\odot}$ away from the centre (second
dredge-up).
The total duration of this phase of inner penetration
of surface convection is 0.55 Myr, after which 
the CNO-burning shell is activated again, and the star
begins the AGB evolution.

\begin{figure}
\caption{The variation with time of 
the temperature at the base of the convective envelope
(top), luminosity due to helium burning (middle), and core mass (bottom)
of a $5M_{\odot}$ FST model during the AGB evolution.
Time was counted from the beginning of the AGB evolution.}
         \label{lumtb}%
\end{figure}

\begin{figure}
\caption{The variation with time of luminosity (top),
mass loss rate (middle) and total mass (bottom)
of the model presented in Fig.~\ref{lumtb}.}
         \label{mlomass}%
\end{figure}

\begin{figure}
\caption{Evolution with time of surface $^3{\rm He}$ (top),
lithium (middle) and fraction of nuclear energy due to
proton-proton reactions (bottom) of the FST 
model presented in Fig.~\ref{lumtb}.}
         \label{litio}%
\end{figure}

\begin{figure}
\caption{Evolution with time of the surface chemical
abundances of the CNO elements (top) and of some
neon, sodium and magnesium isotopes (bottom) of 
the FST model presented in Fig.~\ref{lumtb}.}
         \label{chem}%
\end{figure}

\section{Standard AGB evolution}
Shortly after the second dredge-up the external layers
of the star contract and heat, thus favouring the re-ignition
of the CNO-burning shell. Hereinafter, the CNO cycle will
be the only nuclear source supporting the star, with
the only exception of the contribution of a $3\alpha$
burning shell which once every $\sim 3000$ yr is
ignited in a thermally unstable way, leading to a
thermal pulse (TP).

Fig.~\ref{lumtb}-~\ref{litio} show the temporal evolution of 
some physical quantities related to the evolution 
of the FST model with initial mass $5M_{\odot}$.
We note a rapid increase of the stellar luminosity
(Fig. ~\ref{mlomass}, top panel), which is 
associated with the fast increase of $T_{\rm bce}$
(Fig. ~\ref{lumtb}, top panel), which is
almost doubled after three TPs. We will see that 
the rapidity of this rise is strongly dependent on the
convective model adopted.

The periodic strong drop of the global luminosity and of
$T_{\rm bce}$ marks the ignition of TPs, which trigger an
expansion of the layers above the helium burning shell
with the consequent stop of CNO burning and shrinking
of the surface convective envelope.

The CNO luminosity is generated in a shell which is
$\delta M \sim 10^{-4}M_{\odot}$ wide, and which
progressively moves outwards, burning at higher 
and higher gravities and temperatures: this determines
the increase of luminosity shown in the upper panel
of Fig.~\ref{mlomass}, which is eventually 
halted by the strong mass loss, which reduces the 
mass of the envelope, as can be seen in the bottom
panel of Fig.~\ref{mlomass}. In the same figure we
see the rapid increase of the mass loss with 
luminosity (middle panel): 
within Bl\"{o}cker (1995) prescription, 
and with the adopted value of the free parameter $\eta_R$,
we see that a maximum value of 
$\sim 2\times 10^{-4} M_{\odot}/$yr is attained,
almost in conjunction with the maximum luminosity
of the star.
Therefore, the rapidity with which the
luminosity increases plays a fundamental role
in determining the total duration of the whole
AGB evolution, triggering a fast reduction 
of the mass of the envelope. For the model discussed here, 
we see that the AGB phase lasts $\sim 70,000$ yr.

For a few TPs we note from the bottom panel of 
Fig.~\ref{litio} that a non-negligible 
fraction of the nuclear energy release 
is due to the proton-proton ($p-p$) chain: 
during the 4th interpulse phase this latter 
channel provides $\sim 10\%$ of the luminosity 
of the star. This is due essentially
to $^3{\rm He}$ burning at the base of the external 
envelope (see the top panel of Fig.~\ref{litio}),
which becomes efficient as soon as the 
temperature reaches $\sim 4\times 10^7$K.
The $p-p$ contribution becomes negligible 
as soon as the surface $^3{\rm He}$ 
is all burnt (bottom panel). 
We also note that this is associated with the lithium 
production via the Cameron \& Fowler (1971) mechanism, 
as can be seen by the large increase of the 
surface lithium abundance shown in the middle 
panel of the same figure. The end of
$^3{\rm He}$ burning at the base of the convective envelope
is associated with the temporary plateau in the global
luminosity rising which can be detected in the 
top panel of Fig.~\ref{mlomass}, after 
$\sim 12000$ yr.

We end this general description of AGB evolution
with a glance at the nucleosynthesis at the base
of the convective envelope: we recall that convection
is generally extremely fast in homogenizing the whole
external region; therefore the ejecta of these stars
are practically determined by the changes of the
chemical abundances at the base of the external zone.

We note from the top panel of Fig.~\ref{chem} that 
for the first $\sim 15000$ yr the oxygen abundance
is unchanged, while $^{14}{\rm N}$ is produced at the 
expenses of $^{12}{\rm C}$: this is the signature of a
CN cycle, which is operating at temperatures not
exceeding $\sim 7\times 10^7$K. Later on, 
when $T_{\rm bce}$
exceeds  $\sim 8\times 10^7$K, oxygen is depleted
at the base of the envelope, with the consequent
increase of the $^{14}{\rm N}$ and $^{12}{\rm C}$ abundances. 
After $\sim 15$ TPs, we can cleary see the signature 
of the Third Dredge-Up (TDU) in the strong increase of the
surface $^{12}{\rm C}$ abundance following each TP.
More precisely, we find the first TDU episode
after 13 TPs; the efficiency parameter $\lambda$
(defined as the ratio between the mass which is
dredged up following each TP and the mass gained 
by the H-exhausted region since the precedent TP)
reaches an asymptotic value of $\lambda \sim 0.5$
after 19 TPs.

These results confirm, at least 
within the FST framework,
the possibility of attaining deep oxygen burning
at the base of the external convective zone of these
stars: the final $^{16}{\rm O}$ abundance is $> 10$ times
lower, while the ejecta would have an
average oxygen content a factor of $\sim 3$ lower
than the initial abundance.

From the bottom panel of Fig.~\ref{chem} we see an
early phase of $^{22}{\rm Ne}$ destruction, which favours
a temporary production of sodium, increased by another
factor of $\sim 2$ (in the comparison of the current
sodium abundance with the initial value, we must
recall that the second dredge-up had already favoured
a sodium increase of a factor $\sim 3$). Shortly
after the beginning of oxygen depletion, the 
surface sodium abundance starts decreasing due to
the NeNa chain. 

Later on, when TDU becomes operative, a considerable
amount of $^{22}{\rm Ne}$ is carried outwards to the surface
following each TP: this favours sodium production
again, as we see in the bottom panel of fig~\ref{chem}.
The overall sodium abundance of the ejecta would be
sligtly lower than the initial value.

Finally, we note that $^{24}{\rm Mg}$ is also depleted at
the base of the external convective zone: the final
abundance is lower with respect to the initial
value by a factor of $\sim 20$.

\section{The influence of the convective model}
We stressed in Sect. 3 that prior to the TPs phase 
the three evolutions are very similar.
Here we focus our attention
on the AGB phase, keeping in mind that at the
beginning of this phase the structures of the three 
models are practically identical.

\begin{figure}
\caption{The comparison of the temporal variation of 
the luminosity of three
models with initial mass $M=5M_{\odot}$ calculated with
three different prescriptions for the treatment of the
regions unstable to convection.}
         \label{lumin}%
\end{figure}
 
In Fig.~\ref{lumin} we show the variation with time 
of the luminosity for the three models discussed.
We see a large difference starting already at the first
TPs; the more efficient the convective model is, the more
rapid is the increase of luminosity as the evolution proceeds.
We see that the FST model achieves a maximum luminosity
which is $\sim 70\%$ larger than the corresponding value of
the MLT17 model.

This behaviour can be seen as an extreme case of the
break-down of the core mass vs. luminosity
relation which occurs when the hydrogen burning
region is not detached from the convective region
(Tuchman et al. 1983, Bl\"ocker \& Sch\"onberner
1991).

Such a large difference in the evolution of luminosity 
is inevitably reflected in the mass loss rate: 
we see from Fig.~\ref{mass} a much faster
consumption for the FST model, which loses 
the whole envelope within $\sim 70,000$ yr, to be
compared to the $\sim 130,000$ yr associated to the
MLT21 model, and $\sim 200,000$ yr of the MLT17
model. We therefore can see that the 
duration of the whole AGB phase may
differ by a factor of $\sim 3$ according to the 
convective model adopted.

\begin{figure}
\caption{The variation with time of the total mass
of the models shown in Fig.~\ref{lumin}.}
         \label{mass}%
\end{figure}

Before entering into the details of the consequences
which these differences may have for the chemistry
of the ejecta, hence for the way in which these sources
pollute the interstellar medium, we examine the
reasons for such large differences.

\begin{figure}
\caption{The variation with time of the total
luminosity (top) and of the maximum CNO nuclear
energy release (bottom) for the three models
discussed in Fig.~\ref{lumin}. Only the early
evolution along the AGB is shown.}
         \label{early}%
\end{figure}

We show in Fig.~\ref{early} the comparison 
between the evolution of the three models, 
limited to the first 10,000 yr of AGB, or, 
equivalently, the first 3-4 TPs.
The upper panel shows the total luminosity of the star,
in the bottom panel we report the maximum  
value of the CNO-burning coefficient for nuclear energy 
release. While the latter quantity is
very similar for each of the three models, the luminosities
diverge. We may therefore conclude that while the
internal structure of the models in terms of
thermal stratification and chemical profile
is similar up to the CNO-burning shell, some
difference must arise in the immediate proximity
of the latter, close to the inner boundary of the
convective envelope.

\begin{figure}
\caption{The internal nuclear structure of the 
models presented in Fig.~\ref{lumin} at the
maximum luminosity between the 3rd and 4th
TPs. The abscissa gives the distance 
(in solar masses) from the centre of 
the star, and the ordinate
shows the coefficient for nuclear energy release.
The thin vertical lines indicate the location of
the inner border of the convective envelope.}
         \label{epsnucl}%
\end{figure}

To better clarify this point, we show in
Fig.~\ref{epsnucl} the interior stratifications
of the coefficient for the nuclear energy release;
for all the three models the figure refers to
the point of maximum luminosity during the 
fourth interpulse phase, $\sim 9000$ yr
after the beginning of the AGB evolution.
The mass coordinate of the MLT17 and MLT21
models has been artificially shifted
in order to have the peak value of 
$\epsilon_{\rm nucl}$ at the same abscissa. This
shift is much less than $10^{-3}M_{\odot}$. The thin
vertical lines indicate the 
location of the bottom of the convective envelope
for each model.

\begin{figure*}
\centering{
           }

\caption{Internal distribution of radiative gradient
  (left), and temperature (right) for the same
         models shown in fig~\ref{epsnucl}.}
   \label{superad}
\end{figure*}

For each of the three models we see a very similar
behaviour. In the proximity of the $\epsilon_{\rm nucl}$ peak
there is a secondary maximum 
at the border of the surface convection, which carries
$^3{\rm He}$-rich material into the CNO-burning shell;
during these first TPs, as already discussed in
Sect.4, the $p-p$ contribution is non-negligible
(see the bottom panel of Fig.~\ref{litio}).
Also, we see a secondary maximum $\sim 2\times
10^{-4}M_{\odot}$ away, which is entirely due to 
lithium burning; the contribution of the latter
to the overall energy release, however, is
negligible.

If we compare the three models, we note a
strong similarity up to the peak of the shell
and just beyond, with the only 
difference that the location of the inner
border of the convective external zone is
closer to the CNO peak within the FST model,
while it is a few $10^{-5} M_{\odot}$ further
away in the MLT17 case; the MLT21 model is
intermediate. The proximity of
the convective border to the CNO peak
carries more $^3{\rm He}$ within a region whose
temperature is typical of CNO-burning regions
($T \sim 7\times 10^7$K), thus triggering a 
considerable extra luminosity.
This extra contribution is smaller the more
distant the CNO peak and the base of the
envelope are, and it is practically negligible
within the MLT17 model.

From the above discussion it becomes evident
that in understanding the different
luminosities of the models, despite
the similarity characterizing their structures 
up to the CNO-burning shell, a key role is
played by the exact location of the inner border
of the convective envelope, or, equivalently, its
distance from the peak of the CNO shell. 
Within the framework of the Schwartzschild's
criterion, which is used to fix the neutrality
point according to the condition $\nabla_{\rm rad}=
\nabla_{\rm ad}$, the location of the border is
furhter inward the steeper the
$\nabla_{\rm rad}$ profile towards the external layers. 

The comparison of the variations
of the $\nabla_{\rm rad}$ within the three models
presented in Fig.~\ref{epsnucl} is reported
in the left panel of Fig.~\ref{superad}. 
We see in all cases that $\nabla_{\rm rad}$ 
increases from the peak of the shell, 
because of the rapid decrease of temperature
and pressure and the increase of the opacity;
near the convective region the
FST $\nabla_{\rm rad}$ profile keeps the same slope,
while the MLT models, particularly MLT17,
become progressively flatter, thus moving 
the neutral point, where 
$\nabla_{\rm rad}=\nabla_{\rm ad}$, outwards.

The reason for this behaviour can be found in the
difference in the efficiency of the three convective
models, and the relative effect on the temperature
profile. It is well known that the FST model is more
efficient in high-efficiency convective regions, thus
requiring a lower degree of overadiabaticity to carry
the same amount of energy flux by convection: in
the layers next to the inner border of surface 
convection shown in Fig.~\ref{epsnucl} the FST 
overadiabaticity is $\sim 0.8$ dex smaller.
In a small region very close to the border the
trend is reversed, because the MLT models attain an
extremely low level of overadiabaticy: this is due
to a physical inconsistency of the MLT assumption
that the mixing length is proportional to $H_p$,
which is meaningless near the border,
where the same quantity is expected to vanish, as  
is consistently described in the FST.
\footnote{yet this region
is so efficient convectively that this has no 
consequences for the thermodynamical structure of these
regions, the gradient being very close to the
adiabatic value.}

Within the instability regions, the fraction of 
energy which is carried by radiation decreases
more rapidly in the FST than in 
the MLT models; the necessity of keeping the 
radiative flux $F_{\rm rad}$ higher requires a lower and 
flatter profile of $\nabla_{\rm rad}$
(we recall that $F_{\rm rad} \propto 1/\nabla_{\rm rad}$;
it is the necessity of matching this profile within
the convective region which forces the MLT radiative
gradient profile to bend near the border, thus 
shifting the location of the neutrality point outwards.

\begin{figure*}
\centering{
           }
\caption{The variation of the abundances of some
chemicals at the surface of the star as a function
of the total mass (decreasing due to mass loss). 
Left: CNO abundances; Right: Neon, Sodium and 
Magnesium abundances.}
   \label{abund}
\end{figure*}

The higher efficiency of the FST model thus has two
important consequences:

\begin{itemize}

\item{The FST $\nabla_{\rm rad}$ profile is steeper,
favouring a more internal location of the formal border;
this in turn favours a larger luminosity, because
we have an efficient extra contribution of the
$p-p$ chain.}

\item{The lower value of the overadiabaticity required
in the FST model (and, partially, in the MLT21
model) leads to a smaller temperature gradient
within the instability region
(see the right panel of Fig.~\ref{superad}), 
which, in turn, acts in favour of a larger 
contribution of the extra luminosity supplied by 
$^3{\rm He}$ burning within these stellar layers.}

\end{itemize}

\noindent
From a chemical point of view the differences among 
the three models are also relevant.
Fig.~\ref{abund} shows the variation of the abundances
of some key elements included in our nuclear
network; in the left panel we report the evolution
of the CNO surface abundances, in the right 
$^{20}{\rm Ne}$, $^{23}{\rm Na}$ and $^{24}{\rm Mg}$ are shown.
In this case, to have an idea of the mean chemical
composition of the ejecta of the star, we use the
mass as abscissa instead of time 
(Ventura et al. 2001); the
difference is relevant, because most of the mass
is lost in correspondence of the maximum luminosity,
in a time interval which is very short.

We note in the left panel of Fig.~\ref{abund} 
that all three models achieve oxygen depletion,
though in MLT17 the final depletion is
just by a factor of $\sim 2$. The MLT17 model, 
which yields a longer lifetime, 
experiences more TDUs, thus
it produces more $^{12}{\rm C}$ and $^{14}{\rm N}$. 
As in the FST case, we find that the 
parameter connected to the efficiency of the
TDU approaches an asymptotic value of 
$\lambda \sim 0.5$ after $\sim 20$ TPs.

In the right
panel we note the different behaviour of sodium.
In all three cases we have an early phase of
production (which is due to the combined effects of 
the second dredge-up and to the destruction of 
$^{22}{\rm Ne}$ at the beginning of the AGB evolution),
and a later phase of depletion; in the MLT models,
however, the effect of several TDUs make the
surface content of sodium rise again, and 
reach extraordinarily large values, particularly
within the MLT17 model. We also note a different
degree of $^{24}{\rm Mg}$ depletion in the three cases,
$^{24}{\rm Mg}$ reduction being larger the larger the
efficiency of the convective model.

Table ~\ref{ejecta} summarises the chemical 
composition of the ejecta
for each of the three models discussed. As we may see
there are just two robust conclusions which we can
draw, independently of the convective model adopted:
\begin{itemize}

\item{The fraction (in mass) of helium in the 
expelled material is $Y\sim 0.31$ in all cases: this
can be understood on the basis of the fact that it
occurs essentially during the second dredge-up, 
while the following production during AGB is negligible.}

\item{The lithium content of the ejecta is 
$\log(\epsilon(^7{\rm Li})) \sim 1.9$, slightly less
than the population II average abundance. This is because
the phases of lithium production and destruction
happen at the very beginning of the AGB evolution,
when the differences among the three models are
small.}

\end{itemize}

   \begin{table}
      \caption[]{Chemistry of the ejecta.}
         \label{ejecta}
     $$ 
           \begin{array}{c c c c c c c c}
            \hline
            \noalign{\smallskip}
             Model      & Y(a) & Li7(b) & [C12](c) & [N14](c) & [O16](c) & [Na23](c) & [Mg24](c) \\
            \noalign{\smallskip}
            \hline
            \noalign{\smallskip}
            MLT17 & 0.32 & 2.0 & -0.3 & 1.51 & -0.40 & 0.78 & -0.45     \\
            MLT21 & 0.32 & 1.9 & -0.5 & 1.34 & -0.55 & 0.60 & -0.83     \\
            FST   & 0.31 & 1.9 & -0.7 & 1.13 & -0.60 &-0.16 & -0.95     \\
            \noalign{\smallskip}
            \hline
         \end{array}
     $$ 
\begin{list}{}{}
\item[$^{\mathrm{a}}$] Helium mass fraction of the ejecta.
\item[$^{\mathrm{b}}$] Lithium content of the ejecta on the
scale $\log(N(^7\rm{Li}))=\log(^7\rm{Li}/\rm{H})+12.00$.
\item[$^{\mathrm{c}}$] Average abundances with respect 
to the initial value:  
$[X]=\log(X_{\rm ejecta}/X_{\rm initial})$. 
\end{list}
   \end{table}

\noindent
For all the other elements, the pollution of the
interstellar medium is extremely dependent on the 
convective model:

\begin{itemize}

\item{In the MLT17 case we would expect little
oxygen depletion, and extremely large sodium
and nitrogen enhancement. The sum C+N+O would be
increased by $\sim 0.8$dex, due to the effects of
several TDUs. We would also probably 
expect strong s-process enrichment. The isotopic ratios 
of magnesium $^{25}{\rm Mg}/^{24}{\rm Mg}$ and 
$^{26}{\rm Mg}/^{24}{\rm Mg}$ 
would be slightly less than unity. These results
are consistent with recent AGB models of the
same metallicity presented by Fenner et al. (2004)}.

\item{In the MLT21 model we would 
expect a larger oxygen depletion and
lower sodium enhancement.The increase of 
the sum C+N+O would be lower ($\sim 0.4$dex). The 
isotopic ratios of magnesium would both be
around $\sim 3$.}

\item{The FST model achieves the largest oxygen 
depletion, while the sodium content is slightly
underabundant with respect to the solar value.
The C+N+O abundance is conserved; the isotopic
ratios of magnesium are similar to the MLT21 case.}

\end{itemize}

In summary, we have seen that the stellar yields 
depend dramatically on the convection model. There
are many other uncertainties in the results,
but those are less critical.

We take as an example the sodium content of the ejecta:
we saw that the results change significantly according
to the efficiency of convection at the base of the outer
convective zone. Fig.~\ref{sodium} shows the variation with mass
of the surface sodium abundance within the three models 
discussed so far (full, dotted, and dashed lines), plus a
further FST model (dashed-dotted line) calculated 
by assuming the {\it lower limit} given in Angulo et al. (1999) 
for the reactions destroying sodium 
($^{23}{\rm Na}(p,\gamma)^{24}{\rm Mg}$ and 
$^{23}{\rm Na}(p,\alpha)^{20}{\rm Ne}$)
and the {\it upper limit} for reactions creating sodium
($^{22}{\rm Na}(p,\gamma)^{23}{\rm Mg}$ and 
$^{22}{\rm Ne}(p,\gamma)^{23}{\rm Na}$).
We see that the influence of changing all the cross-sections
involving sodium is far less than the role played by 
convection.

The problems of reaction rates and mass loss will
be discussed in a forthcoming paper.

\section{What implications are there 
for the self-enrichment scenario?}

Deep spectroscopic observations of giant and TO stars
within globular clusters have shown star-to-star
variations in surface chemical composition. 
Particularly, the observations
of NGC 6752 by Gratton et al. (2001) showed the presence of
an O-Na anticorrelation in all stars observed, independently
of their evolutionary stage, including those near the TO. Within 
the same cluster, strong hints of a Mg-Al anticorrelation
were also found. Similar results were obtained for 
other clusters (NGC 6397, M30 and M55: Carretta 2003; 
M71: Ramirez \& Cohen 2002; M5: Ramirez \& Cohen 2003). 
These results have also been recently confirmed
by a deep analysis of the surface abundances of giant 
stars belonging to M3 and M13 by Sneden et al. (2004), where
a clear C-O anticorrelation was found.

The aforementioned results indicate that the surface matter
of these stars has been subject to nuclear processing
though CNO, Na-Ne and Mg-Al chains. Since these
anomalies were observed also in TO stars, where the internal 
temperatures are too low to allow such nuclear
reactions, it was suggested (Ventura et al. 2001; Ventura
et al. 2002) that the stars
displaying such surface chemical anomalies must
have been contaminated by the ejecta
expelled during the AGB evolution of an early generation of 
intermediate mass stars (Cottrell \& Da Costa 1981, 
D'Antona et al. 1983), or must have been formed
by these ejecta (D'Antona et al. 2002).

Recently, this scenario has been questioned by Denissenkov
\& Weiss (2001). Denissenkov \& Herwig (2003), 
based on the computations of a $5M_{\odot}$
model with metallicity $Z=0.0001$, showed that a simultaneous
depletion of oxygen and enhancement of sodium (as indicated
by the anticorrelation observed) is made very difficult
by sodium burning at high temperatures, those necessary
to deplete oxygen; their analysis is independent of 
the amount of extra mixing assumed at the base of 
the convective envelope. 

We postpone to a forthcoming paper
the discussion of the role which IMS may
play in the pollution of the interstellar medium within GCs,
because a more complete analysis, covering the whole range
of masses involved, and the most appropriate metallicities,
rather than a single model, is required.

In any case, the present work clearly suggests that 
among all the uncertainties included
in the input physics adopted to calculate the evolutionary 
sequences the treatment of convection has the greatest impact. 
AGB evolution is one of the few cases in stellar 
astrophysics where the results obtained are 
globally dependent on the convective model 
adopted, in terms of time scale, luminosity, temperature and
nucleosynthesis.

Let us summarize again what we can say about
some of the important elements for the self-enrichment
scenario.

\begin{enumerate}

\item{Lithium:
The constant lithium content of the ejecta, independent of 
the convective model, is in good agreement with the
observed lithium abundances of NGC 6397 TO stars (Bonifacio
et al. 2002), which all show approximately the same
population II standard value.}

\item{Helium:
D'Antona et al. (2002) and D'Antona \& Caloi (2004)
have shown that the high helium content of the ejecta,
confirmed by the present calculations, 
plays a role in the morphology of the horizontal branches 
of GCs.} 

\item{Sodium and Magnesium:
From Fig.~\ref{sodium} we see that the MLT models can achieve
sodium production quite efficiently, because of the TDU carrying
fresh $^{22}{\rm Ne}$ (later converted into sodium) to the surface;
on the other hand, the FST model, living shorter because of
a larger mass loss during the whole AGB evolution,
ejects almost all the envelope mass well before the stage
when TDU can operate significantly. 

In terms of sodium-oxygen anticorrelation, if 
the adopted mass loss rate holds, only the MLT21 model 
seems to pollute the interstellar medium in the ``right'' 
direction, because apart from producing sodium
efficiently it also depletes oxygen (see Table ~\ref{ejecta} and
Fig.~\ref{abund}), though other problems arise, because
in this case the polluted material would have 
magnetisium isotope ratios 
$^{25}{\rm Mg}/^{24}{\rm Mg}$ and 
$^{26}{\rm Mg}/^{24}{\rm Mg}$ 
around $\sim 3$, which disagrees with 
the measured magnesium isotopic ratios by Yong et al.
(2003) for giant stars in NGC 6752. This problem
is shared by the FST model.}

\item{CNO: The C+N+O content of the ejecta is almost
unchanged with respect to the initial value 
in the FST models, which would be in
agreement with the almost constant C+N+O value
found for various GCs stars observed (M92: Pilachowski
et al. (1988); NGC288 and NGC 362: Dickens et al. (1991);
M3 and M13: Smith et al. (1996); M4: Ivans et al. (1999)).
The MLT models lead to a strong increase of the C+N+O
abundance, due to the effects of several TDUs.}

\end{enumerate}

\begin{figure}
\caption{The variation of the surface sodium abundance
with mass within the three models precedently discussed, and 
a further FST model where sodium production is favoured
chosing an appropriate nuclear network.}
         \label{sodium}%
\end{figure}

The self-enrichment hypothesis must be evaluated
on the basis of a much deeper analysis, but it is clear
that the uncertainties connected to convection can no
longer be neglected in the computation of the ejecta of
the various models.

\section{Conclusions}
We examine the effects of changing the treatment of 
convection on the overall evolution of intermediate 
mass stars. We compare the stellar evolution models of 
initial mass $M=5M_{\odot}$ calculated with the MLT
treatment of the instability regions with those 
calculated with the FST model for turbulent 
convection. We show that while the evolutionary 
phases preceding TPs are scarcely
affected (with the exception of RGB and early AGB
effective temperatures), the AGB evolution is heavily influenced
by the convective model adopted. 

A higher convection efficiency at the bottom of the
outer convective zone of AGBs leads to a shorter distance
between the edge of the CNO-burning shell and the base
of external convection; this, in turn, leads to an extra-
luminosity (which, in the first TPs, is entirely due to
$^3{\rm He}$ burning) which triggers a faster growth of the
luminosity itself, hence of mass loss.

From the physical point of view, the overall duration of the
AGB evolution might be shorter by a factor of $\sim 3$
if the MLT treatment of convection is replaced by the FST
model. Noticeable changes can be seen even if, still in
the MLT framework, the free parameter $\alpha$ is changed.
The maximum luminosity achieved by the FST is $\sim 70\%$
larger, and the number of TPs (hence, of third dredge-up
episodes) experienced by the FST model is considerably 
lower.

As for the chemical content of the ejecta, the only robust
conclusion which can be drawn is that these stars
pollute the interstellar medium with material which
is extremely Helium-rich ($Y_{\rm ejecta}=0.31$ in all cases)
and with a lithium content which is only a factor 
of two lower than the standard population II value.

The extent of oxygen depletion, of sodium enrichment and of 
the total C+N+O abundance of the ejecta are strongly
dependent on the convective model adopted.

Convection is one of the most important
parameters in the AGB evolution. Any evaluation 
of the validity of the self-enrichment scenario
for GCs must deal with the uncertainties connected 
to convection before deriving conclusions.


\begin{thebibliography}{}

   \bibitem[1999]{angulo} Angulo, C., et al. 1999, Nucl. Phys. A, 656, 3 
   
   \bibitem[1985]{blo1} Bl\"ocker, T. 1995, A\&A, 297, 727 

   \bibitem[1991]{blo2} Bl\"ocker, T., \& Sch\"onberner, D. 1991, 
      A\&A, 244, L43 

   \bibitem[1958]{bohm} B\"ohm-Vitense, E. 1958, Z. Astroph., 46, 108 

   \bibitem[2002]{bonifacio} Bonifacio, P., Pasquini, L., Spite, F.,
      Bragaglia, A., Carretta, E., Castellani, V. et al. 
      2002, A\&A, 390, 91 

   \bibitem[1971]{cameron} Cameron, A.G.V., \& Fowler, W.A. 1971, 
      ApJ, 164, 111 

   \bibitem[1996]{canuto1} Canuto, V.M., Goldman, I., \& Mazzitelli, 
      I. 1996, ApJ, 473, 570 

   \bibitem[1991]{canuto2} Canuto, V.M., \& Mazzitelli, I. 1991, 
      ApJ, 370, 295 
   
   \bibitem[2003]{carretta} Carretta, E. 2003, Mem. S.A.It. Suppl.,
      3, 90 

   \bibitem[1988]{caughlan} Caughlan, G.R. \& Fowler, W.A. 1988, 
      Atomic Data Nucl. Tab. 40, 283 
   
   \bibitem[1976]{cloutman1} Cloutman, L. \& Eoll, J.G. 1976, ApJ, 
      206, 548
   
   \bibitem[1980]{cloutman2} Cloutman, L. \& Whitaker, R.W. 1980, ApJ, 
      237, 900
      
   \bibitem[1981]{cottrel} Cottrell, P.L. \& Da Costa, G.S. 1981, ApJ, 
      245, 79

   \bibitem[2003]{franca4} D'Antona, F. \& Caloi, V. 2004, ApJ, 
      611, 871

   \bibitem[2002]{franca3} D'Antona, F., Caloi, V., Montalban, J.,
      Ventura, P., \& Gratton, R. 2002, A\&A, 395, 69
   
   \bibitem[1983]{franca2} D'Antona, F., Gratton, R. \& 
      Chieffi, A. 1983, Mem. S.A.It., 54, 173 

   \bibitem[1996]{franca1} D'Antona, F. \& Mazzitelli, I. 1996, 
      ApJ, 470, 1093

   \bibitem[1996]{deng1} Deng, L., Bressan, A., \& Chiosi, C. 1996a, 
      A\&A, 313, 145

   \bibitem[1996]{deng2} Deng, L., Bressan, A., \& Chiosi, C. 1996b, 
      A\&A, 313, 159

   \bibitem[2003]{denis1} Denissenkov, P. \& Herwig, F. 2003, ApJ, 
      590, L99

   \bibitem[2003]{denis2} Denissenkov, P. \& Weiss, A. 2001, ApJ, 
      559, L115

   \bibitem[1991]{dickens} Dickens, R.J.,, Croke, B.F.W., Cannon, R.D.,
      \& Bell, R.A. 1991, Nature, 351, 212
   
   \bibitem[2004]{fenner} Fenner, Y., Campbell, S., Karakas, A.I.,
      Lattanzio, J.C., \& Gibson, B.K. 2004, MNRAS, 353, 789

   \bibitem[1996]{freytag} Freytag, B., Ludwig, H.G., \& Steffen, M. 
      1996, A\&A, 313, 497

   \bibitem[2001]{gratton} Gratton, R., Bonifacio, P., Bragaglia, A. 
      et al. 2001, A\&A, 369, 87

   \bibitem[1996]{grossman1} Grossman, S.A. 1996, MNRAS, 279, 305

   \bibitem[1993]{grossman2} Grossman, S.A., \& Narayan, R. 1993, 
      ApJS, 89, 361 

   \bibitem[2000]{herwig2} Herwig, F. 2000, A\&A, 360, 952

   \bibitem[2004]{herwig3} Herwig, F. 2004, A\&A, 605, 425

   \bibitem[1997]{herwig1} Herwig, F., Bl\"ocker, T., \& Sch\"onberner, D. 
      1997, A\&A, 324, L81

   \bibitem[1988]{iben1} Iben, I.J. 1975, ApJ, 196, 525

   \bibitem[1988]{iben2} Iben, I.J. 1976, ApJ, 208, 165

   \bibitem[1999]{ivans} Ivans, I.I., Sneden, C., Kraft, R.P., et al., 
        1999, AJ, 118, 1273

   \bibitem[1999]{italo} Mazzitelli, I., D'Antona, F., \& 
       Ventura, P. 1999, A\&A, 348, 846
   
   \bibitem[1988]{pilachowski} Pilachowski, C.A. 1988, ApJ, 326, L57

   \bibitem[2002]{ramirez1} Ramirez, S., \& Cohen, J.G.
      2002, AJ, 123, 3277

   \bibitem[2003]{ramirez2} Ramirez, S., \& Cohen, J.G.
      2003, AJ, 125, 224

   \bibitem[1981]{renzini} Renzini, A., \& Voli, M. 
      1981, A\&A, 94, 175

   \bibitem[1991]{sackmann} Sackmann, J., \& Boothroyd, A.I. 
      1991, ApJ, 366, 529

   \bibitem[1965]{schwa1} Schwarzschild, M., \& Harm, R. 
      1965, ApJ, 142, 855

   \bibitem[1999]{schwa2} Schwarzschild, M., \& Harm, R. 
      1967, A\&A, 145, 486

   \bibitem[1996]{smith} Smith, G.H., Shetrone, M.D., Bell, R.A.,
      Churchill, C.W., Briley, M.M. 1996, AJ, 112, 1511

   \bibitem[2004]{sneden} Sneden, C., Kraft, R.P., Guhathakurta, P.,
      Peterson, R.C., Fulbright, J.P. 2004, AJ, 127, 2162
 
   \bibitem[1983]{tuch }Tuchman, Y., Glasner, A., \& Barkat, Z.
      1983, ApJ, 268, 356

   \bibitem[2004]{ventura5} Ventura, P., Castellani, M.
      2004, A\&A, in press

   \bibitem[2001]{ventura1} Ventura, P., D'Antona, F., 
      Mazzitelli, I. \& Gratton, R. 2001, ApJ, 550, L65 

   \bibitem[1998]{ventura2} Ventura, P., D'Antona, F. 
      \& Mazzitelli, I. 2002, A\&A, 393, 215 

   \bibitem[2000]{ventura4} Ventura, P., D'Antona, F. 
      \& Mazzitelli, I. 2000, A\&A, 363, 605 
   
   \bibitem[1998]{ventura3} Ventura, P., Zeppieri, A., D'Antona, F., 
      \& Mazzitelli, I. 1998, A\&A, 334, 953 

   \bibitem[1953]{vitense} Vitense, E. 1953, Zs.Ap., 32, 135 

   \bibitem[1980]{xiong1} Xiong, D.R. 1980, ChA, 4, 234 

   \bibitem[1985]{xiong3} Xiong, D.R. 1985, A\&A, 150, 133 

   \bibitem[1997]{xiong2} Xiong, D.R., Cheng, Q.L., \& Deng, L. 
      1997, ApJS, 108, 529 

   \bibitem[1997]{yong} Yong, D., Grundahl, F., Nissen, P.E., \&  
      Shetrone, M.D. 2003, A\&A, 402, 985 

\end{thebibliography}
\end{document}